\pdfoutput=1
\documentclass[letterpaper,twocolumn,10pt]{article}
\usepackage{usenix}
\usepackage{graphicx}
\usepackage{amsmath,amssymb}
\usepackage{booktabs}
\usepackage{multirow}
\usepackage{cite}
\usepackage{url}
\usepackage{tikz}
\usetikzlibrary{arrows.meta, positioning, shapes.geometric}

\begin{document}

\title{\Large{\bf SoK: Resilience and Fault Tolerance in Cyber-Physical Systems}}

\author{
{\rm Rahul Bulusu}\\
Georgia Institute of Technology\\
\texttt{rbulusu3@gatech.edu}
}
\date{}
\maketitle

\begin{abstract}
Cyber-Physical Systems (CPS) now support critical infrastructure spanning transportation, energy, manufacturing, medical devices, and autonomous robotics. Their defining characteristic is the tight coupling between digital computation and continuous physical dynamics which enables sophisticated autonomy but also creates highly non-linear failure modes. Small disturbances at sensors, firmware, networks, or physical interfaces can propagate through estimation and control pipelines, producing cascading instabilities that defy traditional single-layer reasoning.

This Systematization of Knowledge (SoK) unifies nearly two decades of CPS resilience research into a structured Origin–Layer–Effect (OLE) taxonomy. This taxonomy provides a cross-layer lens for understanding how faults arise, how they propagate, and why unrelated CPS failures often share deep structural similarities. By mapping representative systems including RockDrone, MAYDAY, M2MON, HACMS, Byzantine fault-tolerant control, and learning-based recovery mechanisms onto the taxonomy, we reveal patterns of coverage, persistent blind spots, and recurring pathways of fault amplification.

Our analysis identifies four structural gaps that span multiple CPS domains: (1) physical-model manipulation, (2) ML-enabled control without stability guarantees, (3) semantic inconsistencies between formal models and firmware, and (4) inadequate forensic visibility across cyber and physical layers. These insights motivate new directions for resilient CPS design, integrating robust control, runtime monitoring, formal assurance, and system-level visibility.
\end{abstract}

\section{Introduction}

Cyber-Physical Systems (CPS) now exists in our daily lives, governing domains in which digital decisions directly influence the physical world. Unlike conventional software, CPS mishaps lead not merely to data loss but to real, tangible harm. Drones crash onto crowds, autonomous vehicles misinterpret road conditions, industrial machinery fails catastrophically, and medical devices deliver incorrect dosages. These failures arise because CPS tightly integrate sensing, computation, actuation, and physical dynamics, forming complex feedback loops that amplify small perturbations into full-system failures. Errors that remain inconsequential in traditional computing environments can trigger catastrophic outcomes when they interact with the dynamics of the physical plant.

To illustrate these risks, consider the RockDrone attack \cite{son2015rock}, where a single acoustic tone excites mechanical resonances in a MEMS gyroscope, producing biased angular velocity estimates. These small sensor perturbations propagate through the estimator and into the attitude controller, ultimately destabilizing the UAV. This illustrates the central challenge of CPS resilience: faults originating in one layer can cascade nonlinearly through the entire feedback loop.

The expansion of autonomy has intensified these challenges. Modern CPS depend on heterogeneous sensing modalities, distributed embedded systems, wireless communication networks, cloud-assisted computation, and learning-enabled components for perception and planning. Each of these layers introduces its own assumptions about timing, trust, noise, and error bounds. When these assumptions are violated by environmental disturbances, hardware degradation, or adversarial manipulation, the resulting behavior often escapes the abstractions of any single discipline.

Resilience, in this context, means more than traditional robustness or reliability. A resilient CPS must maintain acceptable behavior under both stochastic disturbances and strategically adversarial actions. It must also provide operators with sufficient visibility to understand why failures occur so that future systems can avoid repeating them. Despite substantial progress across multiple communities like robust control, sensor fusion, embedded systems, security, and formal verification, CPS resilience research remains fragmented.

\subsection{The Need for Systematization}

The research community has produced a rich but fragmented body of work on CPS resilience. Studies examine GPS spoofing \cite{kerns2014uncharted}, LiDAR illusions \cite{shin2017illusion}, acoustic gyroscope manipulation \cite{son2015rock}, CAN bus injection \cite{cho2016error}, timing interference \cite{zhang2001networked}, firmware vulnerabilities, and formal verification of UAV platforms \cite{hacms}. Each provides valuable depth on one part of the resilience problem, yet CPS failures are inherently cross-layer phenomena.

This motivates a Systematization of Knowledge (SoK) that unifies these disparate threads using a structural Origin–Layer–Effect taxonomy. Rather than enumerating attacks or defenses, we map them into this design space, exposing structural patterns and blind spots that are not visible when examining papers individually.

\subsection{Contributions}

This SoK makes three primary contributions.

\textbf{First,} we propose a cross-layer taxonomy for CPS resilience that explicitly factors (1) the \emph{origin} of a disturbance, (2) the \emph{architectural layer} where it manifests, and (3) the \emph{effect type}. This provides a common vocabulary spanning control theory, embedded systems, and security.

\textbf{Second,} we map representative CPS resilience frameworks including MAYDAY \cite{mayday}, M2MON \cite{m2mon}, RockDrone \cite{son2015rock}, Learn2Recover \cite{chatzilygeroudis2019reset}, HACMS \cite{hacms}, and Byzantine fault-tolerant control \cite{bftcontrol} onto this taxonomy. This reveals overlaps, conflicts, and unprotected design regions.

\textbf{Third,} we analyze UAVs, autonomous vehicles, industrial systems, and medical CPS to extract cross-domain structural patterns. Across domains, CPS failures often stem from shared structural weaknesses such as sensor trust assumptions, timing sensitivities, model–firmware mismatches, and insufficient forensic visibility.

\subsection{Scope and Methodology}

We draw from control, embedded systems, CPS security, robotics, and industrial reports. We include work that (i) involves closed-loop CPS dynamics, (ii) captures multi-layer fault propagation, or (iii) proposes defenses informed by physical modeling. We exclude purely IT security and purely control theory unless they materially inform CPS resilience.

We recursively expanded our collection starting from canonical CPS attacks (e.g., GPS spoofing, CAN injection, acoustic sensor attacks) and CPS defense mechanisms (e.g., Simplex, runtime monitoring, robust estimation). Each paper was classified along our taxonomy axes and used to refine the taxonomy.

\section{Background}

Cyber-Physical Systems are defined by closed-loop interactions between the physical world and computational processes. Understanding resilience requires examining the structure of this loop and the foundational principles governing its stability.

\paragraph{Terminology.} Throughout this paper, we distinguish between three 
related but distinct concepts. A \emph{fault} is any deviation in the sensor, actuator, 
or internal software signals. An \emph{error} is a resulting deviation in the 
estimated system state. A \emph{failure} occurs when errors propagate to violate 
a safety- or mission-critical invariant. Adversarial attacks induce faults 
intentionally, while control-semantic faults arise when the physical and 
computational interpretations of system behavior diverge.

\subsection{CPS Architectural Structure}

A CPS typically consists of five interacting subsystems: physical processes, sensors, computational controllers, actuators, and communication networks. The physical process evolves continuously according to natural laws or engineering dynamics. Sensors observe physical variables such as position, velocity, temperature, pressure, or acceleration, but measurements are corrupted by noise, bias, drift, and environmental effects \cite{farrell2008gnss}. Controllers attempt to estimate the system state and compute actuator commands that achieve a control objective, such as stabilizing a drone, tracking a lane, or maintaining pressure within a safe band. Actuators apply forces or changes that alter the physical world. Communication channels relay sensor readings, actuator commands, and state estimates, often under bandwidth or latency constraints and sometimes over insecure protocols like CAN or Modbus \cite{cho2016error}.

In theory, CPS stability emerges from carefully designed feedback loops in which controllers are synthesized using accurate models of plant dynamics and well-characterized disturbance profiles. In practice, real-world deployments deviate from these idealized assumptions. Noise is rarely Gaussian, dynamics is nonlinear and time-varying, communication is unreliable, and environmental conditions change over seasons, weather patterns, and operational cycles. These deviations compromise theoretical guarantees and open windows for adversarial interference.

\subsection{Control-Theoretic Principles}

Control theory provides mathematical guarantees about the behavior of dynamic systems. Stability ensures that errors decay over time or remain bounded under bounded disturbances; controllability ensures that the system can reach desirable states; observability ensures that internal states can be inferred from measurements. Classical designs including PID controllers, linear–quadratic regulators (LQR), and model predictive control (MPC) rely on accurate models of the plant. When model fidelity degrades, even small perturbations can destabilize an otherwise stable system.

State estimation techniques like Kalman filtering assume specific noise distributions and often linearity. In adversarial or heavily degraded environments, these assumptions fail. Attackers have exploited these assumptions by injecting crafted perturbations that appear statistically consistent with expected noise \cite{teixeira2012attack}. Meanwhile, timing-related issues including jitter, communication delay, asynchronous sampling and others, introduce new instability modes not captured by standard continuous-time analysis \cite{zhang2001networked}. Networked control theory has extended classic results to handle some of these effects but often still presumes disturbances.

\subsection{CPS Complexity/Challenges in Real Deployments}
Real-world CPS are non-stationary: their dynamics evolve due to mechanical wear, environmental change, or configuration drift. UAVs experience changes in inertia from payloads or battery depletion; autonomous vehicles encounter unpredictable road conditions, varying friction coefficients, and changing lighting; industrial plants undergo gradual degradation of pumps, valves, and sensors. CPS also increasingly incorporate machine learning to enhance perception or high-level planning. Yet ML components lack the stability and interpretability of traditional controllers and are highly sensitive to distribution shifts and adversarial perturbations \cite{cao2019adversarial}. These factors combine to produce systems whose behavior cannot be fully characterized by static models or traditional verification methods, necessitating runtime monitoring and adaptive resilience strategies.

Beyond environmental non-stationary, CPS complexity is intensified by the tight integration of heterogeneous components with fundamentally different design philosophies. Control engineers typically assume well-defined noise bounds, bounded delay, and smoothly varying system dynamics. Embedded systems engineers reason about worst-case execution times, interrupt latency, memory safety, and branching corner cases. Security researchers generally assume that adversaries behave strategically and may violate trust assumptions in arbitrary ways. In many CPS failures, the root cause is not a single flaw in any one layer but rather an unforeseen interaction between these perspectives. A controller may be provably stable under the assumed sampling rate, yet an operating-system task overrun may silently increase latency beyond that threshold. An estimator may filter sensor noise effectively under Gaussian assumptions, yet adversarial perturbations can be shaped to mimic those distributions. These inconsistencies illustrate why CPS cannot be understood purely through isolated disciplinary lenses; resilience requires an integrated systems perspective.

Furthermore, increasing reliance on interconnected services introduces cloud dependencies, remote telemetry, and over-the-air (OTA) updates that fundamentally change the CPS attack surface. When decision-making is partially offloaded to remote servers or when firmware is updated dynamically in the field, trust boundaries shift in ways that complicate verification and runtime confidence. Even phenomena such as network congestion or unanticipated update interactions can create cascading effects that propagate through the physical domain. This entanglement between cyber infrastructure and physical behavior suggests that CPS resilience must incorporate not only local control robustness but also distributed system reliability, secure update channels, and systemic risk analysis.

\section{System Model and Taxonomy of CPS Resilience}

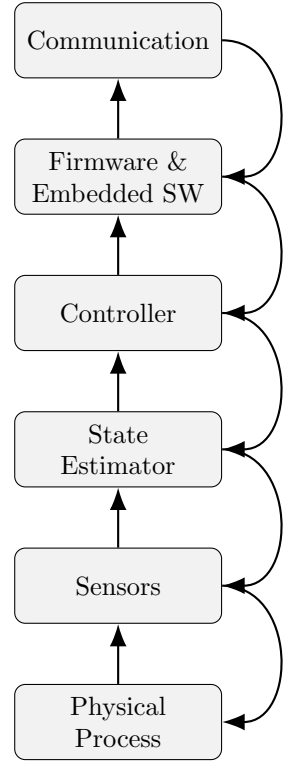
\begin{figure}[t]
\centering
\begin{tikzpicture}[
    node distance=8mm and 12mm,
    box/.style={
        rectangle,
        draw,
        rounded corners,
        align=center,
        minimum width=2.75cm,
        minimum height=1.0cm,
        fill=gray!10
    },
    arrow/.style={-{Latex[length=3mm]}, thick},
]

\node[box] (physical) {Physical\\Process};
\node[box, above=of physical] (sensors) {Sensors};
\node[box, above=of sensors] (estimator) {State\\Estimator};
\node[box, above=of estimator] (controller) {Controller};
\node[box, above=of controller] (firmware) {Firmware \&\\Embedded SW};
\node[box, above=of firmware] (network) {Communication};

\draw[arrow] (physical) -- (sensors);
\draw[arrow] (sensors) -- (estimator);
\draw[arrow] (estimator) -- (controller);
\draw[arrow] (controller) -- (firmware);
\draw[arrow] (firmware) -- (network);

\draw[arrow] (network.east) .. controls +(1,0) and +(1,0) .. (firmware.east);
\draw[arrow] (firmware.east) .. controls +(1,0) and +(1,0) .. (controller.east);
\draw[arrow] (controller.east) .. controls +(1,0) and +(1,0) .. (estimator.east);
\draw[arrow] (estimator.east) .. controls +(1,0) and +(1,0) .. (sensors.east);
\draw[arrow] (sensors.east) .. controls +(1,0) and +(1,0) .. (physical.east);

\end{tikzpicture}
\caption{Conceptual CPS taxonomy diagram illustrating the layered structure of physical processes, sensing, estimation, control, firmware, and communication. Faults and adversarial perturbations may enter at any layer and propagate upward or downward through feedback loops. \textit{Note: Physical dynamics influence all upper layers and information flows up; commands flow down.}}
\label{fig:taxonomy-diagram}
\end{figure}

\begin{table*}[t]
\centering
\begin{tabular}{lll}
\toprule
\textbf{Axis} & \textbf{Values} & \textbf{Examples} \\
\midrule
Origin &
Unintentional &
Sensor noise, hardware wear \\
&
Adversarial &
GPS spoofing, CAN injection \\
&
Control-semantic &
Payload change, miscalibration \\
\midrule
Layer &
Physical process &
Load manipulation, friction change \\
&
Sensors &
LiDAR ghosts, acoustic resonance \\
&
Estimator &
Filter divergence, biased state \\
&
Controller &
Gain tampering, mode errors \\
&
Firmware &
Fault injection, parameter flips \\
&
Network &
Replay, delay, packet drops \\
\midrule
Effect &
Value corruption &
False readings, corrupted commands \\
type &
Timing perturbation &
Jitter, delay, reordering \\
&
Model mismatch &
Outdated dynamics, wrong gains \\
\bottomrule
\end{tabular}
\caption{Summary of our CPS resilience taxonomy. Each disturbance is classified by its origin, layer of manifestation, and effect type.}
\label{tab:taxonomy-summary}
\end{table*}

To reason systematically about resilience and fault tolerance, we introduce a simple abstract system model and a taxonomy for classifying faults, attacks, and defenses. The model is not tied to any specific domain but captures common structure across UAVs, vehicles, and industrial systems.

We consider a discrete-time CPS with state $x_k$, control input $u_k$, disturbance $w_k$, and measurement $y_k$. The plant dynamics are modeled as
\[
x_{k+1} = f(x_k, u_k, w_k),
\]
and the measurement function as
\[
y_k = h(x_k, v_k),
\]
where $w_k$ and $v_k$ represent process and measurement disturbances, respectively. A controller implements a policy
\[
u_k = \pi(\hat{x}_k, r_k),
\]
where $\hat{x}_k$ is an estimate of the state and $r_k$ is a reference signal or high-level command. The estimator computes $\hat{x}_k$ from past inputs and outputs, typically using a model of $f$ and $h$. Network effects introduce delays and packet loss into the feedback loop, while firmware and software layers realize $\pi$ and the estimation algorithm on concrete hardware.

Within this model, faults and attacks can affect any of the functions $f$, $h$, or $\pi$, any of the signals $w_k$, $v_k$, $u_k$, or $y_k$, and the timing with which information flows. We classify these disturbances along three orthogonal axes:
\begin{itemize}
    \item \textbf{Origin} $O \in \{\text{unintentional}, \allowbreak \text{adversarial}, \allowbreak \text{control-semantic}\}$, distinguishing natural noise and wear from strategically crafted perturbations and from mismatches between models and reality.
    
    \item \textbf{Layer} $L \in \{\text{physical process}, \allowbreak \text{sensors}, \allowbreak \text{estimator}, \allowbreak \text{controller}, \allowbreak \text{firmware}, \allowbreak \text{network}\}$, capturing where the disturbance first becomes visible or actionable in the CPS stack.
    
    \item \textbf{Effect type} $E \in \{\text{value corruption}, \allowbreak \text{timing perturbation}, \allowbreak \text{model mismatch}\}$, describing whether it primarily alters signal values, temporal behavior, or the semantic relationship between models and the physical plant.
\end{itemize}
Each concrete fault or attack can be represented as a point $(O,L,E)$ in this space. Resilience mechanisms can then be characterized by which regions of this space they monitor, tolerate, or actively mitigate. For example, a Byzantine-resilient estimator focuses on adversarial value corruptions at the sensor layer, while instruction-level monitors address adversarial value and timing perturbations at the firmware layer.

This abstract view underscores why no single defense mechanism suffices. A robust controller that tolerates bounded noise may remain vulnerable to timing attacks; a runtime monitor that enforces control-flow integrity does not prevent physical spoofing of sensors; a sensor-fusion anomaly detector may miss attacks that manipulate dynamics rather than immediate measurements. Effective CPS resilience requires coverage across multiple regions of the taxonomy.

While the abstract model offers a clean conceptual mapping, real CPS often deviate from these formulations in structured but important ways. For example, hybrid systems combine discrete state machines with continuous dynamics, meaning that faults may cause discrete mode transitions rather than continuous drift. Such mode changes can alter stability margins dramatically, pushing systems from safe operational regions to unstable ones. Similarly, network-induced variability imposes stochastic delays that transform the system into a partially observed Markov decision process, where control actions depend on probabilistic beliefs rather than deterministic estimates. Resilience mechanisms must therefore reason about both continuous uncertainty and discrete transitions, often requiring hybrid verification tools or runtime observers that track the CPS mode as well as its state. Table~\ref{tab:taxonomy-summary} summarizes our taxonomy axes, providing concrete examples for each category. 

\paragraph{Interpretation.} Figure~\ref{fig:taxonomy-diagram} highlights the 
bi-directional nature of CPS fault propagation. Disturbances introduced at the 
physical or sensing layers flow upward into estimation and control, while 
firmware- or network-level timing perturbations propagate downward into actuator 
behavior. Timing faults go both directions. This structure explains the 
recurrence of certain attack patterns across domains and why single-layer 
defenses often fail to contain multi-layer cascades.

To illustrate these interactions, consider the conceptual multi-layer model shown in Figure~\ref{fig:taxonomy-diagram}. At the bottom, physical processes generate inherently noisy and uncertain dynamics. Above them, sensors create an observational interface that is susceptible to drift and spoofing. Estimators convert these observations into approximate internal states, which feed into controllers that compute actuation signals. Finally, firmware and hardware layers operationalize these decisions with their own timing and resource constraints. In this diagram, faults propagate upward or downward depending on their origin, and attacks can enter at any layer. A comprehensive taxonomy must therefore capture the bi-directional influence between layers: physical disturbances reshape software behavior, while software faults induce nontrivial physical effects. This multi-directional coupling is what makes CPS resilience uniquely challenging compared to traditional computing systems.

\section{Fault Models}

Understanding resilience requires classifying the types of faults that can arise. We distinguish unintentional faults, adversarial faults, and control-semantic faults, each with distinct characteristics and propagation pathways.

\subsection{Unintentional Faults}

Unintentional faults arise from natural sources such as sensor noise, drift, hardware degradation, and environmental fluctuations. Sensor drift introduces slow bias accumulation, which may go unnoticed until controllers saturate or exhibit unstable oscillations. Noise becomes especially problematic in multi-sensor fusion systems where high-frequency disturbances propagate through estimators. Environmental disturbances including wind gusts, vibrations, electromagnetic interference, produce conditions that deviate from the assumptions used during controller design. In practice, engineers often design controllers with margins intended to tolerate such deviations, but it is rarely feasible to anticipate the full range of conditions encountered over a system’s lifetime.

Timing and synchronization faults pose serious challenges. Clock drift causes misalignment between sensors and controllers, leading to inconsistent estimates. Network-induced latency results in control actions based on outdated information, degrading performance or destabilizing the system \cite{zhang2001networked}. In distributed CPS, inconsistent clocks can cause different subsystems to make mutually incompatible decisions. Mechanical wear creates nonlinear dynamics that invalidate linearized models used for control. Valves may stick, motors may lose torque, and mechanical backlash may introduce dead zones or other things.

\subsection{Adversarial Faults}

Adversarial faults differ from natural disturbances in that attackers intentionally craft perturbations to induce unsafe behavior. GPS spoofing misleads navigation systems without modifying internal software \cite{kerns2014uncharted}. LiDAR spoofing creates ghost obstacles or deletes real ones, undermining autonomous vehicle perception \cite{shin2017illusion}. Acoustic attacks on UAV gyroscopes exploit mechanical resonance phenomena \cite{son2015rock}. These attacks operate entirely at the physical layer, bypassing traditional cyber-security defenses.

Communication-based adversarial faults exploit insecure protocols. CAN bus injection allows attackers to send malicious steering or braking commands \cite{cho2016error}. Replay attacks on Modbus mimic legitimate sensor updates. Timing manipulation exploits the sensitivity of controllers to delayed or reordered messages. Firmware-level adversarial faults arise when attackers modify embedded software. Fault-injection attacks exploit voltage or clock glitches to cause microcontrollers to skip instructions or bypass safety checks \cite{tupsamudre2018survey}. These adversarial faults are often percise and intermittent, making them hard to distinguish from disturbances using purely statistical detectors.

In taxonomy terms, these attacks primarily occupy the adversarial--sensor and 
adversarial--firmware regions, with value-corruption effects dominating. GPS 
spoofing corresponds to $(O{=}adv, L{=}sensor, E{=}value)$, while CAN injection 
maps to $(O{=}adv, L{=}network, E{=}value)$ and timing attacks to 
$(O{=}adv, L{=}network, E{=}timing)$.

\subsection{Control-Semantic Faults}

Control-semantic faults occur when physical and computational interpretations diverge. These faults are also precise because they often arise without explicit software bugs. Payload changes modify drone inertia, requiring new PID gains. If firmware does not update these gains, the UAV becomes unstable, as demonstrated in MAYDAY’s crash analyses \cite{mayday}. Calibration drift causes multi-modal fusion systems to misinterpret sensor data even when each modality functions correctly. Discretization and quantization errors accumulate in implementations of continuous-time controllers. Seemingly harmless changes, such as altering sampling frequency, can move a system from stable to unstable regimes.

These semantic mismatches often account for a large portion of real-world CPS failures. They are particularly insidious because they frequently evade pre-deployment testing. Systems may pass laboratory tests under nominal conditions yet fail in the field when confronted with combinations of model drift, timing deviations, and operating regimes not represented during design.

\subsection{Mathematical Formulation of Attacks}
We can formalize the fault types discussed above by augmenting the standard linear time-invariant (LTI) system model. Consider the nominal system:
\begin{align}
    x_{k+1} &= A x_k + B u_k + w_k \\
    y_k &= C x_k + v_k
\end{align}
where $A, B, C$ represent the system, input, and output matrices. We categorize adversarial injections as additive vectors affecting specific signal paths.

\paragraph{Sensor Attacks.}
A sensor attack injects a vector $\delta^y_k \in \mathbb{R}^p$ into the measurement equation:
\begin{equation}
    y^a_k = C x_k + v_k + \delta^y_k
\end{equation}
In a \textit{Naive Spoofing} attack, $\delta^y_k$ is arbitrary. In a \textit{Stealthy False Data Injection (FDI)} attack, the adversary designs $\delta^y_k$ to remain within the null space of the anomaly detector. For a standard $\chi^2$ failure detector comparing the residual $r_k = y_k - C\hat{x}_{k|k-1}$, an attack is stealthy if the induced residual shift $\Delta r_k$ satisfies:
\begin{equation}
    \| \Delta r_k \| \leq \tau
\end{equation}
where $\tau$ is the detection threshold. This requires the attacker to possess knowledge of the system matrices $A$ and $C$ to simulate the estimator's internal state.

\paragraph{Actuator and Network Attacks.}
Attacks on the control signal (e.g., CAN injection) can be modeled as an additive disturbance on the input:
\begin{equation}
    x_{k+1} = A x_k + B (u_k + \delta^u_k) + w_k
\end{equation}
Unlike sensor attacks, actuator attacks directly alter the physical energy of the system. A \textit{Replay Attack} captures a valid sequence of inputs $u_{t:t+N}$ and measurements $y_{t:t+N}$ and replays them at a later time $k$. This is particularly effective against systems without physical challenge-response authentication, as the replayed signals satisfy the dynamic constraints $y \approx h(x,u)$ despite decoupling the controller from the current physical reality.

\paragraph{Dynamics/Parametric Attacks.}
The most precise attacks involve altering the physics itself (e.g., changing a drone's payload or friction coefficients), effectively changing the system matrix $A$ to $A'$.
\begin{equation}
    x_{k+1} = A' x_k + B u_k + w_k
\end{equation}
If the controller continues to calculate inputs $u_k = -K \hat{x}_k$ based on the assumption of matrix $A$, the closed-loop eigenvalues of $(A' - BK)$ may shift into the unstable region (outside the unit circle) without triggering standard value-checking monitors until critical instability is reached.

\section{Attack Surfaces in CPS}

\begin{figure*}[t]
\centering

\begin{tabular}{c c c c c}
\framebox[3.3cm]{Sensors} &
\framebox[3.3cm]{Estimator} &
\framebox[3.3cm]{Controller} &
\framebox[3.3cm]{Firmware} &
\framebox[3.3cm]{Network} \\
[1.2em]

\framebox[3.3cm]{Acoustic resonance} &
\framebox[3.3cm]{Filter divergence} &
\framebox[3.3cm]{Gain tampering} &
\framebox[3.3cm]{Instruction skips} &
\framebox[3.3cm]{CAN spoofing} \\
[1.2em]

\framebox[3.3cm]{GPS spoofing} &
\framebox[3.3cm]{Fusion inconsistency} &
\framebox[3.3cm]{Timing attacks} &
\framebox[3.3cm]{Memory corruption} &
\framebox[3.3cm]{Replay attacks} \\
[1.2em]

\framebox[3.3cm]{LiDAR ghosts} &
\framebox[3.3cm]{State bias} &
\framebox[3.3cm]{Model mismatch} &
\framebox[3.3cm]{Parameter injection} &
\framebox[3.3cm]{MITM interference} \\
\end{tabular}

\caption{CPS Attack Surface Matrix. Rows represent attack types; columns represent architectural layers. Many attacks propagate cross-layer.}
\label{fig:cps-attack-surface}
\end{figure*}

CPS expose a wide range of attack surfaces due to their cross-layer architecture. Figure~\ref{fig:cps-attack-surface} shows that attacks cluster most densely in 
the sensor and firmware columns, with relatively few computation-originating 
attacks that manifest physically. This asymmetry reflects long-standing research 
biases and is a key structural insight of this SoK.
Attacks at one layer propagate through the system, demonstrating the interconnectedness of CPS vulnerability.

\subsection{Sensor-Level Attacks}

Sensors fundamentally trust the physical world. Attackers exploit this trust using physics-based techniques. GPS spoofers emit counterfeit satellite signals to misguide navigation systems \cite{kerns2014uncharted}. LiDAR spoofers use carefully timed laser pulses to inject fake depth points \cite{shin2017illusion}. Camera-based attacks rely on adversarial patterns or flickering lights to confuse vision systems and cause misclassification \cite{cao2019adversarial}. Acoustic interference can exploit gyroscope resonance to introduce measurement drift \cite{son2015rock}. These attacks require no digital compromise and often appear indistinguishable from environmental anomalies, complicating detection.

\subsection{Communication-Level Attacks}

Communication channels represent high-value targets due to their central role in CPS coordination. CAN bus remains widely used in automotive systems despite lacking message authentication. Attackers can inject arbitrary control commands or spoof messages to override legitimate signals \cite{cho2016error}. Industrial protocols like Modbus and DNP3 are equally insecure, enabling replay or man-in-the-middle attacks. Timing-based attacks manipulate the scheduling of messages, inducing very fine instabilities that controllers interpret as natural disturbances \cite{zhang2001networked}. These attacks exploit the fact that many controllers were designed under assumptions of bounded latency.

\subsection{Controller-Level and Firmware Attacks}

Controller implementations and embedded firmware often contain undocumented features, misconfigurations, or insufficient validation. Specification-based monitoring frameworks show that CPS firmware frequently deviates from expected execution patterns, reflecting latent vulnerabilities \cite{bartocci2018specification}. Fault injection through voltage or electromagnetic pulses allows attackers to induce unpredictable state transitions or bypass safety routines \cite{tupsamudre2018survey}. Because firmware is often closed-source and proprietary, it is difficult for third parties to assess its correctness or robustness.

\subsection{Physical-Process-Level Attacks}

Attacks that manipulate the physical process itself represent one of the most crafty threat classes. Stuxnet altered centrifuge speeds while feeding operators falsified sensor data, exploiting the mismatch between physical behavior and control-system assumptions \cite{stuxnet}. BlackEnergy similarly exploited operator trust in industrial HMIs during the Ukrainian power grid attacks \cite{blackenergy}. Adversaries may also manipulate loads, environmental conditions, or mechanical configurations in ways that cause control algorithms to fail gracefully on paper but dangerously in reality. These attacks demonstrate that security mechanisms focused solely on cyber layers are insufficient.

\section{Resilience Techniques}

CPS resilience encompasses detection, robust control, redundancy, runtime monitoring, recovery, and formal assurance. We organize these techniques into thematic categories and relate them to the taxonomy introduced earlier.

\subsection{Detection and Diagnosis Mechanisms}

Detection mechanisms attempt to identify anomalies before they destabilize the system or cause unacceptable degradation. MAYDAY’s cross-domain crash analysis correlates physical trajectories with firmware logs to identify root causes of UAV failures and to distinguish between pilot error, environmental disturbances, and control-semantic bugs \cite{mayday}. M2MON enforces I/O-level invariants that detect inconsistencies between commanded and observed behavior at the firmware boundary \cite{m2mon}. Specification-based monitoring inspects instruction-level execution, providing fine-grained detection of firmware tampering or unexpected control-flow paths \cite{bartocci2018specification}. Statistical anomaly detectors flag deviations in network traffic patterns, such as suspicious CAN messages that do not match learned models of legitimate behavior \cite{cho2016error}. Machine learning–based anomaly detectors identify inconsistencies across multi-modal perception pipelines \cite{cao2019adversarial}. These mechanisms differ in granularity, overhead, and coverage; together, they form a layered defense strategy that can catch both cyber and physical anomalies.

\subsection{Robust and Fault-Tolerant Control}

Robust controllers explicitly account for bounded uncertainty in the model. They maintain stability by designing control laws that tolerate disturbances or parameter variations, often by maximizing margins or using $H_\infty$ techniques. Adaptive control dynamically adjusts controller parameters in real time to compensate for unexpected changes in system dynamics, such as payload variation or actuator degradation. Observer-based techniques detect inconsistencies between predicted and measured states, enabling online fault identification and isolation. Fault-tolerant control architectures reweight sensor inputs or switch to redundancy when faults are detected, maintaining estimation and control performance even when some sensors are compromised or failed \cite{farrell2008gnss}. Byzantine fault-tolerant controllers treat arbitrary sensor failures as adversarial behaviors and design control laws that remain stable despite a subset of malicious or arbitrarily faulty sensors \cite{bftcontrol}.

\subsection{Redundancy and Diversity}

Redundancy increases resilience by providing multiple independent sources of information or control. Hardware redundancy includes duplicate IMUs, redundant power systems, or backup actuators. Modal redundancy combines sensors with different physical principles, such as vision, LiDAR, radar, and inertial sensing \cite{cao2019adversarial}. Algorithmic diversity employs multiple estimation or control strategies in parallel, relying on majority voting or inconsistency detection when different algorithms disagree. Diversity makes it harder for attackers to compromise all modalities simultaneously and reduces the risk that a single modeling error will affect all subsystems.

\subsection{Runtime Monitoring and Enforcement}

Runtime enforcement complements detection by ensuring that system execution adheres to defined safety constraints in real time. Specification-based monitoring validates execution against expected control flows, halting or flagging suspicious deviations and providing an enforcement hook inside the firmware \cite{bartocci2018specification}. M2MON ensures that actuator commands remain consistent with safe operational envelopes and with physical sensor readings \cite{m2mon}. PokeMon enforces state-machine transitions at runtime, preventing the system from entering unsafe or undefined states \cite{pokemonsm}. These monitors are most effective when tied to formally specified invariants or contracts, but even heuristic invariants can improve resilience in practice. A central challenge is managing the computational overhead of monitoring while satisfying the tight real-time constraints imposed by CPS.

\subsection{Architectural Resilience: Simplex and Rejuvenation}
When prevention mechanisms fail and a controller is compromised, the system architecture must ensure physical safety and recover control. Research in this domain focuses on three primary paradigms: trust splitting, delayed input sharing, and continuous rejuvenation.

\textbf{The Simplex Architecture (BlueBox):} 
A fundamental challenge in CPS is that high-performance controllers are complex and prone to bugs, while verifiable controllers are too simple for advanced maneuvering. BlueBox \cite{bluebox} resolves this using a \textit{Simplex Architecture} \cite{sha1999simplex}. It runs an untrusted, high-performance controller (Complex Core) alongside a formally verified safety controller (Safety Core) on isolated hardware. The Safety Core monitors the physical state of the vehicle; if the Complex Core commands an action that would violate a physical invariant (e.g., flipping a drone or exiting a geofence), the Safety Core vetoes the command and switches to a safe fail-over logic.

While widely influential, Simplex \cite{sha1999simplex} provides protection primarily at the controller 
and firmware layers. It does not address sensor spoofing or timing faults, and 
thus covers only a narrow region of our taxonomy.

\textbf{Delayed Input Sharing (BFT++):} 
Traditional Byzantine Fault Tolerance (BFT) is often too slow for real-time control loops. BFT++ \cite{bftplusplus} introduces a specialized primary-backup architecture tailored for CPS inertia. It employs a \textit{Delayed Input Sharing} mechanism where the primary controller processes sensor input at time $t$, while the backup controller receives the same input via a FIFO queue with a delay of $\Delta t$. If a malicious input causes the primary controller to crash or enter an undefined state, the system detects the failure before the backup processes that same input. The backup can then discard the "toxic" input from its queue and assume control. This approach leverages the physical inertia of the system to ride out the momentary gap in control during the handover.

\textbf{Continuous Rejuvenation (YOLO):} 
Advanced persistent threats can hide in controller memory indefinitely. The YOLO framework (``You Only Live Once'') \cite{yolo} counters this by enforcing \textit{Continuous Rejuvenation}. It aggressively resets and reboots controller nodes (e.g., every 100ms) to wipe out resident malware or corrupted state. By rotating active control among redundant nodes, the system ensures that at least one node is always fresh and uncompromised. This technique relies on the physical plant's inertia to maintain stability during the brief reset windows.

\textbf{Learning-Based Recovery:} 
While the above methods address system availability, semantic faults (e.g., actuator degradation) require adaptive logic. Frameworks like Learn2Recover \cite{chatzilygeroudis2019reset} employ reinforcement learning to update control policies in real-time. Unlike static robust control, L2R treats recovery as a regression problem, learning to map current (possibly degraded) actuator capabilities to necessary control inputs to maintain stable flight. Within our taxonomy, Learn2Recover occupies the unintentional/control-semantic 
origin at the controller layer with a model-mismatch effect, representing a 
distinct region from classical adversarial defenses.

\subsection{Formal Methods and Verified CPS}

Formal verification efforts, exemplified by programs like HACMS \cite{hacms}, aim to provide mathematical guarantees about CPS controllers and software components. These techniques use theorem proving, model checking, and code generation to ensure that implementations satisfy high-level safety properties. While promising, they face several limitations. Formal models are necessarily abstractions of reality and may omit important physical phenomena. There is also a semantic gap between verified models and deployed firmware binaries, which tools like M2MON begin to address \cite{m2mon}. Verification does not eliminate the need for runtime monitoring and resilience to unmodeled disturbances.

\subsection{Physical Fingerprinting and Source Identification}
A critical challenge in legacy CPS (particularly automotive CAN bus) is the lack of cryptographic authentication. To address this, researchers have developed physical fingerprinting techniques that authenticate the \textit{hardware} sender rather than the digital credential.

\textbf{Clock Skew (CIDS):} Electronic Control Units (ECUs) rely on crystal oscillators that exhibit microscopic, unique frequency deviations due to manufacturing variations. CIDS \cite{cho2016fingerprinting} measures the inter-arrival time of periodic CAN messages to derive a clock skew ``fingerprint.'' If a compromised ECU attempts to impersonate another node (Masquerade Attack), the discrepancy between the claimed ID's expected skew and the attacker's actual skew reveals the intrusion.

\textbf{Voltage Signatures (Viden):} As attackers evolved to emulate clock skew (``Cloaking'' attacks), defenses shifted to lower-level physical characteristics. Viden \cite{cho2017viden} analyzes the transient voltage characteristics of the CAN bus signal during the transmission of the ACK bit. Since voltage output is determined by the specific physical circuitry and wire impedance of the sender, it provides a robust physical signature that is difficult for an attacker to spoof via software alone.

\section{Case Studies}

\begin{table*}[t]
\centering
\begin{tabular}{lcccccc}
\toprule
\textbf{System} & \textbf{Layer} & \textbf{Fault Type} & \textbf{Detection} & 
\textbf{Mitigation} & \textbf{Cross-Layer} & \textbf{Overhead} \\
\midrule
MAYDAY        & Controller/Physical & Semantic                 & Yes & Partial  & Strong   & Low \\
M2MON         & Firmware I/O        & Adversarial              & Yes & Moderate & Low       \\
Spec. Monitoring & Firmware Execution  & Adversarial              & Yes & Partial  & Weak      & Medium \\
RockDrone     & Sensor Physics      & Adversarial              & Yes & No       & Weak      & Low \\
PyCRA         & Camera Fusion       & Adversarial              & Yes & No       & Weak      & Medium \\
HACMS         & Model/Software      & Semantic                 & Yes & Limited  & Strong   & High \\
Learn2Recover & Controller/Physical & Semantic/Noise           & Yes & Yes      & Moderate & Medium \\
BFT Control   & Sensors/Estimator   & Adversarial              & Yes & Strong   & Strong   & High \\
\bottomrule
\end{tabular}
\caption{Comparison of major CPS resilience frameworks across layers, fault types, capabilities, and overhead.}
\label{tab:cps-resilience}
\end{table*}

\begin{table*}[t]
\centering
\begin{tabular}{lccc}
\toprule
\textbf{System} & \textbf{Dominant Origin $O$} & \textbf{Primary Layer $L$} & \textbf{Primary Effect Type $E$} \\
\midrule
MAYDAY \cite{mayday} &
Control-semantic &
Controller / Physical &
Model mismatch \\
M2MON \cite{m2mon} &
Adversarial &
Firmware I/O &
Value corruption \\
Spec. Monitoring \cite{bartocci2018specification} &
Adversarial &
Firmware execution &
Value + timing perturbation \\
RockDrone \cite{son2015rock} &
Adversarial &
Sensors (gyroscope) &
Value corruption \\
PyCRA \cite{pycra} &
Adversarial &
Sensors (camera / fusion) &
Value corruption \\
HACMS \cite{hacms} &
Control-semantic &
Model / Software &
Model mismatch \\
Learn2Recover \cite{chatzilygeroudis2019reset} &
Unintentional + Control-semantic &
Controller / Physical &
Model mismatch, value correction \\
BFT Control \cite{bftcontrol} &
Adversarial &
Sensors / Estimator &
Value corruption \\
\bottomrule
\end{tabular}
\caption{Mapping representative CPS resilience systems to our taxonomy. Each system primarily targets a subset of origin--layer--effect space, revealing coverage and gaps.}
\label{tab:taxonomy-mapping}
\end{table*}

\begin{figure*}[t]
\centering
\small

\resizebox{\textwidth}{!}{
\begin{tabular}{lccc}
\toprule
\textbf{Region of Taxonomy} &
\textbf{Representative Systems} &
\textbf{Coverage Density} &
\textbf{Notes} \\
\midrule

Adversarial -- Sensor -- Value &
RockDrone, LiDAR spoofing, GPS spoofing, PyCRA &
High &
Most-studied region; strong focus on spoofing attacks. \\

Adversarial -- Network -- Value/Timing &
CAN injection, MITM, replay, NCS timing attacks &
Moderate &
Timing perturbations less explored than value corruption. \\

Adversarial -- Firmware -- Value/Timing &
Spec. Monitoring, Fault injection &
Moderate &
Strong mechanisms exist, but semantic coverage is limited. \\

Unintentional -- Sensor/Physical -- Model mismatch &
MAYDAY, Learn2Recover &
Low--Moderate &
Few works treat model drift as a primary failure mode. \\

Control-semantic -- Controller -- Model mismatch &
HACMS, Simplex, robust control &
Moderate &
Formal guarantees exist but rarely extend to full stacks. \\

Physical-process -- All origins -- Dynamics manipulation &
Stuxnet, load manipulation studies &
Low &
Under-explored; large blind spot across domains. \\

\bottomrule
\end{tabular}
} 

\caption{Coverage of the Origin--Layer--Effect taxonomy by representative CPS resilience systems. Darker regions indicate dense research activity, while lighter rows identify structural blind spots.}
\label{fig:taxonomy-coverage}
\end{figure*}

\subsection{Unmanned Aerial Vehicles}

UAVs exhibit vulnerabilities across sensing, communication, control, and firmware layers. RockDrone’s acoustic gyroscope attack demonstrates how physics-based interference can perturb control loops without digital compromise \cite{son2015rock}. GPS spoofing disorients navigation systems, causing drones to drift or land in unauthorized locations \cite{kerns2014uncharted}. MAYDAY’s forensic framework links flight anomalies to control-semantic inconsistencies and provides a structured approach for post-crash analysis that integrates telemetry, firmware logs, and physical dynamics \cite{mayday}. Learn2Recover provides an example of resilience by allowing UAVs to learn corrective behaviors from previous failures and disturbances \cite{chatzilygeroudis2019reset}. Together, these studies illustrate not only how UAVs fail, but how domain-specific knowledge can be used to design resilience mechanisms that integrate model-based and data-driven approaches.

In taxonomy terms, RockDrone maps to $(adv,\ sensor,\ value)$, GPS spoofing to 
$(adv,\ sensor,\ value)$, and 
MAYDAY to $(control\text{-}semantic,\ controller,\ model)$. This mapping shows 
that UAV incidents, while diverse, occupy consistent regions of the taxonomy.

\subsection{Autonomous Vehicles}

Autonomous vehicles depend heavily on multi-modal perception, which makes them vulnerable to sensor spoofing and fusion inconsistencies. LiDAR spoofing introduces false obstacles; camera attacks confuse visual classifiers; radar jamming disrupts speed estimation \cite{shin2017illusion,cao2019adversarial}. CAN bus injection allows attackers to override vehicle control signals, leading to unintended acceleration or steering changes \cite{cho2016error}. Timing interference in sensor fusion pipelines destabilizes vehicle trajectories when delay patterns violate assumptions baked into control design. Defensive strategies combine sensor fusion consistency checks, redundancy across modalities, anomaly detection leveraging temporal and spatial correlations, and runtime monitoring at both network and controller layers. However, ensuring robust performance under adversarial conditions remains a major research challenge. In particular, vehicles that rely extensively on deep neural networks must cope with adversarial examples and distribution shifts that classical robust control theory does not address.

\subsection{Industrial Control Systems}

Industrial CPS operate in high-consequence environments where downtime is costly and safety constraints are strict. Stuxnet highlighted the dangers of semantic manipulation by altering physical behavior while spoofing sensor data to operators, thereby exploiting trust in human–machine interfaces \cite{stuxnet}. BlackEnergy and related malware demonstrated that power grid operations could be disrupted not only through malware on workstations but also by manipulating control logic and operator displays \cite{blackenergy}. Replay attacks and logic manipulation remain prevalent threats. Despite advances in monitoring and incident response, many industrial controllers still lack basic authentication or integrity mechanisms due to legacy constraints. Upgrading such infrastructure without disrupting operations is a slow process, making resilience an ongoing concern.

\subsection{Medical and Assistive CPS}

Medical CPS, such as infusion pumps, implantable devices, and surgical robots, operate directly on or inside the human body. Faults in these systems have immediate and personal consequences. Timing errors in infusion pumps can lead to under- or over-dosing; sensor failures in closed-loop insulin delivery systems may cause hypoglycemia or hyperglycemia. Adversarial work in this space has shown the feasibility of wireless manipulation of implantable medical devices, although detailed physical-model attacks remain less explored than in UAVs or vehicles. Resilience research here emphasizes fail-safe design, continuous self-checks, and conservative safety margins. These systems illustrate a different point on the resilience trade-off: they often prefer false positives in detection and conservative shutdown behavior over performance degradation.

Table~\ref{tab:taxonomy-mapping} summarizes how these representative systems populate our taxonomy. Notably, most mechanisms concentrate on adversarial value corruptions at the sensor, estimator, and firmware layers, with relatively little attention to timing perturbations or to control-semantic faults that arise from slowly evolving model mismatches.

\subsection{The Challenge of Stealthy Stuxnet-style Attacks}
The Stuxnet attack on Iranian centrifuges serves as the canonical example of a \textit{Replay} and \textit{Integrity} attack working in tandem. The malware recorded valid sensor data (normal centrifuge rotation speeds) for a period of 21 seconds. It then replayed this recorded data to the operator's Human-Machine Interface (HMI) while simultaneously sending malicious commands to the Variable Frequency Drives (VFDs) to spin the rotors to destruction frequencies.

In our taxonomy, this represents a coordinated multi-point failure:
\begin{itemize}
    \item \textbf{Origin:} Adversarial (State-Actor).
    \item \textbf{Layer:} Simultaneous manipulation of Sensor (replay) and Actuator (frequency injection).
    \item \textbf{Effect:} Timing perturbation (delayed data) masking Value corruption.
\end{itemize}
This highlights a critical blind spot in classical monitoring: the "Loop of Fiction." If the monitor resides at the SCADA layer (cyber), it sees the replayed "normal" data. Only a monitor with physical access to the motor back-EMF or acoustic vibrations (physical layer) could have detected the divergence between the reported and actual states.

\subsection{Comparative Analysis}
To synthesize the capabilities of these diverse systems, Table~\ref{tab:cps-resilience} provides a comparative overview of the primary resilience frameworks discussed. It highlights their specific defense layers, mitigation capabilities, and relative overhead. While detection is common across all frameworks, automated mitigation remains rare, particularly for attacks originating at the physical layer. Furthermore, Table~\ref{tab:taxonomy-mapping} details how these systems populate our specific $(O, L, E)$ taxonomy, revealing that most research concentrates on adversarial value corruptions, often leaving timing perturbations and control-semantic model mismatches under-addressed.

\subsection{Cross-Domain Patterns and Lessons}

Across these domains, three cross-cutting patterns emerge when viewed through our taxonomy. First, \emph{most exploitable faults originate at the sensor and firmware layers but manifest as control-semantic failures at the physical layer}. RockDrone and LiDAR spoofing attacks begin as adversarial sensor value corruptions, yet ultimately destabilize attitude or trajectory control. CAN injection similarly originates in network manipulation but results in model-mismatch behavior that looks indistinguishable from misconfiguration or pilot error.

Second, \emph{timing perturbations are systematically under-addressed}. UAV and automotive controllers are designed under assumptions about worst-case latency and sampling rates, yet few deployed defenses explicitly monitor or enforce these assumptions at runtime. Timing attacks on networked control systems \cite{zhang2001networked} and industrial protocols demonstrate that violating temporal contracts can be as damaging as corrupting values, but the literature on timing-aware resilience remains sparse compared to value-centric anomaly detection.

Third, \emph{ML-enabled components amplify both resilience and fragility}. In autonomous vehicles and emerging medical CPS, learning-based perception and decision making increase tolerance to natural variability but introduce new sensitivity to adversarial perturbations and distribution shifts \cite{cao2019adversarial}. Our taxonomy highlights that these components primarily sit at the boundary between sensing and estimation layers, where their failures rapidly propagate into controller decisions. Yet most existing resilience mechanisms treat ML components as black boxes, rather than reasoning about their uncertainty or failure modes in closed-loop settings.

These patterns suggest that domain-specific incidents in UAVs, vehicles, ICS, and medical CPS are manifestations of the same underlying structural issues in CPS resilience, rather than isolated anomalies.

In taxonomy terms, LiDAR spoofing occupies $(adv,\ sensor,\ value)$, CAN 
injection occupies $(adv,\ network,\ value)$, and timing interference maps to 
$(adv,\ network,\ timing)$. Vehicle incidents commonly progress from sensing 
failures to estimator drift to unsafe controller outputs.

\section{Cross-Layer Analysis}

A defining feature of CPS failures is cross-layer propagation. A disturbance originating at one layer often cascades across others. For instance, LiDAR spoofing alters perception, leading to incorrect state estimation, which then triggers faulty actuator commands that destabilize the vehicle. Firmware vulnerabilities allow modifications that subtly alter control gains, interacting with physical models in ways not captured by formal verification, as MAYDAY showed \cite{mayday}. Timing attacks distort the synchrony between sensing, estimation, and actuation, degrading controller performance and reducing robustness margins. From the perspective of our abstract model, such attacks modify both the effective dynamics $f$ and the observation function $h$ in ways unanticipated during design.

Cross-layer resilience requires integrating multiple defense mechanisms. Redundancy mitigates sensor-level attacks but must be combined with anomaly detection to identify inconsistent modalities. Runtime monitors ensure firmware integrity but require alignment with model-level invariants to identify semantic inconsistencies. Adaptive controllers restore stability but rely on accurate state estimates, which in turn depend on trustworthy sensors and reliable communication. End-to-end assurance therefore cannot be achieved by focusing on a single layer. Instead, designers must identify critical dependencies across layers and design defenses that complement rather than duplicate each other.

A deeper examination of cross-layer propagation reveals mathematical structures that explain why attacks at one layer manifest as complex failures elsewhere. In linearized systems, small sensor injection attacks can be expressed as additive disturbances on the observation function $h$. However, when passed through nonlinear estimators or fused with other modalities, these disturbances become multiplicative or state-dependent, leading to estimation bias that grows nonlinearly with system velocity, acceleration, or actuator saturation. This phenomenon is well documented in UAV crash analyses where slight gyroscope drift, once filtered through complementary or Kalman filters, amplifies into significant attitude estimation errors that overwhelm PID control loops. A similar pattern appears in automotive systems: small CAN message delays, when 
propagated through sensor fusion and model predictive controllers, produce 
nonlinear bias growth that can exceed actuation limits during emergency maneuvers.
Such effects show that resilience must account for the \emph{amplification dynamics} of CPS pipelines, not merely their individual errors.

Cross-layer behavior also exhibits what can be described as ``semantic interference,'' where assumptions in one subsystem implicitly constrain or conflict with assumptions in another. For example, a controller may assume synchrony between sensor measurements and actuation, yet the embedded scheduler may reorder these events under high computational load. Similarly, vision-based perception may assume smooth motion between frames, but network or firmware delays can violate such smoothness, causing machine-learning classifiers to misinterpret motion cues. These inconsistencies underscore that CPS correctness is a global property: local correctness does not guarantee global safety when subsystems are tightly interdependent.

Finally, cross-layer analysis reveals that resilience mechanisms often interact in unexpected ways. Robust controllers may mask the symptoms of failing sensors long enough for faults to spread, delaying detection and complicating diagnosis. Runtime monitors may enforce safety conditions but inadvertently induce additional latency that tightens control-loop margins. Learning-based detection techniques may adapt to and inadvertently normalize slow adversarial drift. These interactions emphasize the need for holistic resilience strategies that explicitly model the dependencies and feedback loops connecting each layer of the CPS stack.

\section{Structural Gaps and Open Challenges}

\subsection{Structural Gaps in the Literature}

Our taxonomy and cross-domain analysis reveal several structural gaps that persist across CPS resilience research.

\paragraph{Gap 1: Physical-model manipulation.}
Most defenses assume that plant dynamics remain within a known envelope and treat deviations as noise or drift. In practice, adversaries can deliberately manipulate loads, friction, or environmental conditions to move the system into poorly modeled regimes while remaining within sensor tolerances. Existing work on robust control and attack-resilient estimation \cite{teixeira2012attack,bftcontrol} partially addresses this problem, but few deployed systems monitor for or reason explicitly about \emph{model-regime changes} as a first-class security signal. This gap aligns with the sparsely populated lower-left region of 
Figure~\ref{fig:cps-attack-surface}, where physical-process manipulations 
interact with estimation and control.

\paragraph{Gap 2: Learning-enabled controllers without stability guarantees.}
ML-based perception and planning components are increasingly embedded in closed-loop control, yet most work on adversarial ML ignores the feedback-loop consequences of misclassifications. Conversely, most control-theoretic work assumes fixed, interpretable controllers. There is little support today for end-to-end guarantees that combine neural network verification with nonlinear plant dynamics and runtime uncertainty estimates.

\paragraph{Gap 3: Weak links between models and firmware.}
Formal methods efforts such as HACMS \cite{hacms} verify models and high-level code, while firmware monitoring systems like M2MON \cite{m2mon} and specification-based monitors \cite{bartocci2018specification} enforce properties on binaries. However, few approaches provide a continuous chain of evidence connecting high-level specifications, automatically generated code, hand-written firmware, and the behavior of the compiled binary on real hardware. This semantic gap makes it difficult to know whether verified properties truly hold in the field.

\paragraph{Gap 4: Limited forensic visibility and operator-centric tools.}
Systems like MAYDAY \cite{mayday} illustrate the value of cross-domain crash analysis, yet most CPS deployments still lack the logging, synchronization, and visualization needed to reconstruct failures across cyber and physical layers. Operator interfaces typically expose raw sensor values and alarms, rather than structured explanations in terms of the CPS state, violated assumptions, or taxonomy categories. This lack of visibility slows incident response and prevents organizations from learning systematically from failures.

\subsection{Open Research Challenges}
Despite recent advances, CPS resilience faces several long-term challenges. A major challenge concerns the detection of physical-model manipulation. Adversaries can subtly alter dynamics to push systems into unstable regions while remaining within expected physical ranges, making detection difficult. Robust and adaptive controllers can mitigate some of these effects but are not a cure. Developing methods that distinguish between configuration changes and adversarial manipulations remains an open problem.

A second challenge is verifying the correctness and robustness of machine-learning components. ML models lack explicit, human-readable specifications, making it difficult to understand failure modes or guarantee stability when they are embedded in closed-loop control architectures \cite{cao2019adversarial}. Formal verification of neural networks has made progress in bounding behavior for small input perturbations, but extending these results to end-to-end CPS that include nonlinear dynamics and environmental uncertainty remains a significant research frontier.

A third challenge relates to runtime enforcement. Instruction-level monitors impose overhead, and fine-grained monitoring becomes increasingly expensive as CPS grow in complexity and performance demands. Techniques like M2MON \cite{m2mon} and specification-based monitoring \cite{bartocci2018specification} show promise but require further optimization and careful integration with control-loop timing budgets. Designers must find ways to provide strong enforcement without undermining the very real-time guarantees that make CPS viable.

Finally, bridging the semantic gap between design-time models and runtime firmware remains unresolved. Systems such as HACMS highlight the difficulty of achieving end-to-end verification when real-world systems incorporate legacy code, proprietary firmware, and unmodeled physical dynamics \cite{hacms}. Ensuring consistency between control models, software implementations, and real-world behavior will require new tools that combine static analysis, dynamic monitoring, and physically grounded testing.

Another emerging challenge lies in the interplay between human operators and CPS autonomy. As systems grow more intelligent and complex, human oversight becomes both more important and more difficult. Operators may misinterpret system alerts due to overwhelming information or poorly designed interfaces, as seen in industrial incidents where human–machine interfaces presented misleading operational states. Building resilience therefore requires designing interfaces that expose internal CPS confidence, uncertainty, and anomaly explanations in interpretable ways. The integration of explainable AI for perception and decision-making components may become essential to ensuring that operators can meaningfully supervise autonomous behavior.

Lastly, resilience must confront the reality that CPS increasingly participate in networked ecosystems rather than isolated deployments. Autonomous vehicles interact with roadside units, cloud-based maps, and other vehicles; drones communicate with ground stations; industrial plants rely on remote monitoring. These interdependencies create systemic risk where local failures propagate across the network. Ensuring resilience at scale thus requires combining CPS-specific robustness with distributed-systems fault tolerance, including consensus protocols, authenticated sensing networks, and resilient coordination algorithms. Long-term research must bridge CPS theory with network security and distributed computing to provide guarantees that hold not only for individual systems, but for entire interconnected infrastructures.

\section{Conclusion}

This SoK provides a comprehensive, cross-layer synthesis of resilience and fault tolerance in Cyber-Physical Systems. By examining faults, attack surfaces, detection methods, robust control strategies, runtime monitors, and reconfiguration mechanisms, we build a holistic understanding of how CPS fail and how they can be strengthened. Our systematization reveals recurring vulnerabilities rooted in model assumptions, timing sensitivities, sensor trust, and physical-layer interactions. It also highlights how promising defenses, from robust and adaptive control to runtime monitoring and learning-based recovery, can begin to address these vulnerabilities when combined thoughtfully.

Future CPS will increasingly integrate learning-enabled perception, distributed 
coordination, and highly networked physical infrastructure. Achieving resilience 
under these conditions requires defensive architectures that span sensors, 
estimation pipelines, controllers, firmware, and human operators. Our taxonomy 
and cross-layer analysis provide a structural foundation for these efforts, 
highlighting the propagation pathways that must be monitored and the semantic 
assumptions that must be validated end-to-end.

\bibliographystyle{plain}
\bibliography{references}

\end{document}